\documentclass[usenatbib,usegraphicx]{mn2e}

\newcommand{\msun}{\mathrm{M}_\odot}
\newcommand{\rsun}{\mathrm{R}_\odot}

\newcommand{\mtwodot}{\dot{M}_2}

\newcommand{\vul}{V407~Vul}
\newcommand{\rxj}{RX~J0806+1527}

\title[Period Changes in Ultra-compact Double White Dwarfs]{Period
  Changes in Ultra-compact Double White Dwarfs}

\author[T. R. Marsh, G. Nelemans]
       {T. R. Marsh$^1$, G. Nelemans$^2$\\
        1. Department of Physics, University of Warwick, Coventry CV4 7AL, UK\\
        2. Department of Astrophysics, Radboud University Nijmegen, Nijmegen,
        The Netherlands\\
}

\date{Accepted ;
      Received ;
      in original form}

\pagerange{\pageref{firstpage}--\pageref{lastpage}}
\pubyear{2005}

\begin{document}

\maketitle

\label{firstpage}

\begin{abstract}
  In recent years there has been much interest in the nature of two stars,
  \vul\ and \rxj, which are widely thought to be binary white dwarfs of very
  short orbital period, 570 and 321 seconds respectively. As such they should be
  strong sources of gravitational waves and possible ancestors of the accreting
  AM~CVn stars. Monitoring at X-ray and optical wavelengths has established
  that the period of each star is decreasing, at rates compatible
  with that expected from gravitational radiation. This has been taken to
  support the ``unipolar inductor'' model in which the white dwarfs are detached
  and the X-rays produced by the dissipation of magnetically-induced electric
  currents. In this paper we show that this interpretation is incorrect because
  it ignores associated torques which transfer angular momentum between the
  spin of the magnetic white dwarf and the orbit. We show that this torque is
  $\sim 10^5$ times larger than the GR term in the case of \vul\, and $\sim 10$ times
  larger for \rxj. For \vul, the unipolar inductor model can only survive if
  the white dwarf spins $\sim 100$ times faster than the orbit. Since this could
  only come about through accretion, the validity of the unipolar inductor
  appears questionable for this star. We also consider whether accretion models can fit the
  observed spin-up, concluding that they can, provided that a mechanism exists
  for driving the mass transfer rate away from its equilibrium value.
\end{abstract}


\section{Introduction}

Over the past decade, observations have established the existence of a
population of some 100--200 million double white dwarfs within our
Galaxy \citep{Marsh:friends,Napiwotzki:SPY}. These and their
descendants are thought likely to be a dominant source of low
frequency gravitational waves in the Galaxy
(e.g. \citealt{Hils:GWR,Nelemans:GWR0}), and are a possible progenitor population
of Type~Ia supernovae. A significant fraction of these binary stars
are close enough that gravitational wave losses will cause them to
undergo mass transfer within a Hubble time. Most will merge to
form single stars, variously suggested to be Type~Ia supernovae, or in
the majority of cases, sdB, sdO or R~CrB stars (e.g. \citealt{Webbink:DDs,
Iben:ddrcrb, Saio:merger}). If any systems survive the onset of mass
transfer as binary stars, then they would become semi-detached
accreting binary stars with white dwarf donors, properties matched by
the AM~CVn stars, which feature helium-dominated spectra and orbital
periods which range from 10 to 65 minutes.

The question of survival as a binary is key to whether double white
dwarfs are the ancestors of AM~CVn stars and is important to the
prediction of gravitational waves. The nearest we can get to a proof
that the start of mass transfer can be survived is to identify
accreting pairs of double white dwarfs with periods short of 10
minutes since there are alternative routes for systems with periods
longer than this \citep{Nelemans:AMCVn, Podsiadlowski:AMCVn}. In this
context, two stars, \vul\ and \rxj\ have generated much interest in
the past few years because they show periods of 570 and 321 seconds
respectively, and there are several reasons to think that these may be
orbital \citep{Cropper:V407Vul}, rather than, for example, the spin of a magnetic accreting
white dwarf as seen in ``intermediate polars'' (but see \citealt{Norton:V407Vul}).
The key features of these systems are (i) X-ray light curves which are
off for about half the period \citep{Cropper:V407Vul,Israel:RXJ0806_0}, 
suggestive of a spot of emission on a spinning star, (ii) optical and infrared light curves that show the
same period and no other \citep{Ramsay:V407Vul,Ramsay:RXJ0806,Israel:RXJ0806}, 
(iii) weak optical line emission \citep{Ramsay:V407Vul2,Israel:RXJ0806}, and 
(iv) very soft X-ray spectra. Items ii--iv in particular count against
\cite{Norton:V407Vul}'s intermediate polar model.

Even on the assumption that we are seeing orbital periods, questions remain over
the nature of these stars.  There are currently three popular models which are
(a) the polar model in which we see X-rays from the magnetic poles of a white
dwarf locked to its companion \citep{Cropper:V407Vul}, (b) the unipolar inductor
(UI) model in which X-rays comes from the dissipation of electric currents generated
from a slight asynchronism between the spin of a magnetic white dwarf and its
companion \citep{Wu:V407Vul}, and (c) the direct impact model in which X-rays
come from the point at which the mass transfer stream impacts the white dwarf,
which can happen directly in these very compact systems
\citep{Nelemans:AMCVn,Marsh:V407Vul,Ramsay:V407Vul2}. The polar and direct
impact models rely on accretion and so imply that the systems have survived
contact; the system in the UI model is not in contact and
therefore has not necessarily had to survive it.  The polar model has fallen
somewhat out of favour because of the weak optical line emission, an absence of
or, at best, weak polarisation
\citep{Ramsay:V407Vul2,Reinsch:RXJ0806,Israel:RXJ0806_2} and the X-ray spectra,
which are unusually soft, even for polars (but see
\citealt{Cropper:ultracompacts}). The two most promising models left are the
UI and direct impact models. It was soon realised that period
changes might help distinguish between these models since on the detached
UI model it was predicted that the period would decrease while in
accreting models the period should increase as the donor expands. In a series of
papers, the periods of both \vul\ and \rxj\ have been definitively established
to be decreasing
\citep{Strohmayer:V407Vul,Hakala:RXJ0806,Strohmayer:RXJ0806_1,Israel:RXJ0806_2,Hakala:RXJ0806_1,Ramsay:V407Vul3,Strohmayer:RXJ0806_2},
which, in spite of some problems which face the UI model \citep{Marsh:V407Vul,Barros:V407Vul}, 
has been taken to be strong evidence in its favour \citep{Cropper:ultracompacts}.

In much of this work it has been assumed, almost by default, that, since the
systems in the unipolar inductor (UI) model are detached, their orbital periods
change at a rate governed by gravitational radiation alone (see e.g.
\citealt{Strohmayer:V407Vul,Cropper:ultracompacts,Hakala:RXJ0806_1,Willes:electric}). In this paper
we show that this is very far from being the case, and that on the contrary the
observed period changes rule out \cite{Wu:V407Vul}'s original model
once-and-for-all. We further show that it can only survive in a very different
form which necessarily requires accretion to have occurred, raising the question
of why it ever should have ceased. We begin by discussing the period change
expected for the UI model.

\section{Period changes from induction torques}
\label{sec:induction}
The key point about the UI model which has not always been appreciated is that
although the two white dwarfs are detached, their period evolution reflects 
the combination of angular momentum loss through gravitational radiation 
\emph{and} induction-driven angular momentum interchange within the
system.  Although this was stated explicitly by \cite{Wu:V407Vul}, who also 
present equations which correctly account for the effect, its importance seems not to
have been generally recognised. As we will show, in \cite{Wu:V407Vul}'s model,
the angular momentum interchange term dwarfs the GR loss term, and therefore
the idea that the UI model evolves in the same way as a pair of detached stars
is completely wrong.

In the UI model the X-rays come from the dissipation of
electric currents generated by magnetic induction. This implies that a
torque $T$ exists which transfers angular momentum between the spin
of the magnetic primary star and the binary orbit. If the spin angular
frequency of the primary star is $\Omega_s$, then the rate of work
done by the torque is $T \Omega_s$ (we define the sign of $T$ such
that a positive value spins the primary star up). An equal but opposite
torque acts to extract angular momentum from the orbit, removing
energy from it at rate $T \Omega_o$ where $\Omega_o$ is the orbital
angular frequency.  The difference between the two is dissipated, at
least in part, in the form of X-rays so that
\begin{equation}
L_X \le T (\Omega_o - \Omega_s). \label{eq:torque}
\end{equation}
Therefore the rate of change of orbital angular momentum $J$ is given by
\begin{equation}
\dot{J} = \dot{J}_{GR} - \frac{L_X}{(1-\alpha) \Omega_o}, \label{eq:jdot}
\end{equation}
where we assume from now on that equation~\ref{eq:torque} is an
equality and where we have introduced $\alpha = \Omega_s/\Omega_o$,
following \cite{Wu:V407Vul}. In this equation $\dot{J}_{GR}$ is the rate of
angular momentum lost due to gravitational waves alone. For two
detached stars, the rate of angular momentum loss immediately leads to
the period change through $\dot{P}/P = 3 \dot{J}/J$,
therefore whether the period change occurs at the GR rate depends upon
the relative magnitudes of the two terms in Eq.~\ref{eq:jdot}. If we
multiply through by the orbital angular frequency, then we obtain
an illuminating version:
\begin{equation}
\dot{E} = - L_{GR} - \frac{L_X}{(1-\alpha)}, \label{eq:edot}
\end{equation}
where $L_{GR}$ is the luminosity in gravitational waves and $\dot{E}$ is the
rate of change of orbital energy.
Equation~\ref{eq:edot} is identical in physical content to Equation~9 of
\cite{Wu:V407Vul}, except that for simplicity we have ignored the moment of inertia of the
secondary star as it makes only a small difference to the results. Equation~\ref{eq:edot}
shows that the simple GR formula for the period derivative can only be used if 
\begin{equation}
L_X \ll \left|1-\alpha\right| L_{GR} .
\end{equation}
We now show that, on \cite{Wu:V407Vul}'s model, this is not at all the case for
either \vul\ or \rxj.

\subsection{\vul\ and \rxj}
\label{sec:lgr}
In order to estimate the gravitational wave luminosities, let's 
assume that we \emph{are} seeing the pure GR rate, in which case it can be
shown that
\begin{equation}
L_{GR} = \frac{5 c^5 \dot{P}^2}{1152 G \pi^2},
\end{equation}
independent of the masses. The observed rates of period change of \vul\ and \rxj\ are
$\dot{P} = 3 \times 10^{-12} \,\mathrm{s/s}$ \citep{Strohmayer2004,Ramsay:V407Vul3} and
$3.7 \times 10^{-11}\,\mathrm{s/s}$ \citep{Israel:RXJ0806_2,Strohmayer:RXJ0806_2}
respectively. These give $L_{GR} = 1.4 \times 10^{33}$ and $2.1 \times 10^{35}$ ergs/s
for the two systems.

The X-ray luminosity of \vul\ is $\sim 10^{35} (d/1\,\mathrm{kpc})^2$~ergs/s
\citep{Ramsay:V407Vul3}.  From Eq.~\ref{eq:edot}, and assuming
\cite{Wu:V407Vul}'s value of $1-\alpha = 0.001$, one would therefore predict
that the induction torque term is $\sim 10^5 (d/1\,\mathrm{kpc})^2$ times larger
than the GR term.  In other words the observed rate of period change in \vul\ is
\emph{far too small} compared to a prediction based upon \cite{Wu:V407Vul}'s
model. The model is nearer the mark in the case of \rxj\ for which $L_X
\sim 2 \times 10^{33} (d/1\,\mathrm{kpc})^2$~ergs/s \citep{Strohmayer:RXJ0806_2},
and the predicted induction torque term is ``only'' $\sim 10$ times the GR term. However,
this still shows that it is incorrect to assume that period changes in the UI
model run at a rate given solely by GR.

\begin{figure}
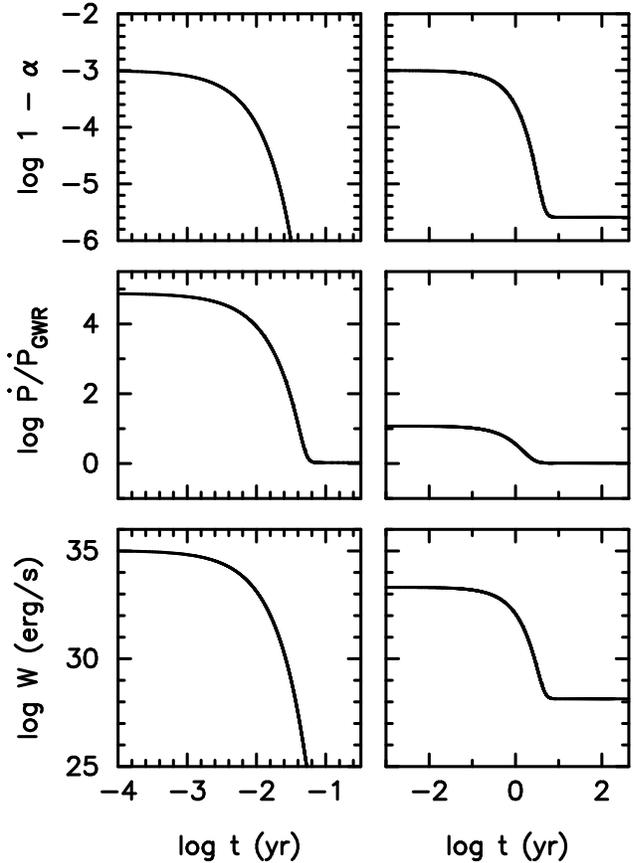

\hspace*{\fill}
\includegraphics[angle=270,totalheight=11.5cm]{Fig1_left.ps}
\hspace*{\fill}
\includegraphics[bb= 78 194 591 344, angle=270, totalheight=11.5cm, clip=]{Fig1_right.ps}
\hspace*{\fill}
\caption{Synchronisation in the unipolar inductor model. From top to bottom the
panels show the degree of asynchronism, the period derivative scaled by the
GR-only value and power dissipated as a function of time. The left-hand panel
starts with $1-\alpha = 0.001$ and $L_X = 10^{35}\,$erg/s to match \vul\ while 
the right-hand panel similarly matches \rxj. Note that the horizontal axes of
the two panels differ from each other.}
\label{fig:sync}
\end{figure}

A problem for the UI model has always been its short lifetime before
synchronisation occurs, estimated to be $\sim 1000$ years by \cite{Wu:V407Vul}.
In order to have (at least) two such systems within a kiloparsec of the Sun,
this suggests a formation rate of order 1 per year within the Galaxy, unless the
asynchronism can be regenerated by some as-yet-undetermined mechanism
\citep{Cropper:ultracompacts}. We will now show that the problem is actually
much worse than even this estimate suggests. During synchronisation, the spin
rate of the primary will change much faster than that of the orbit, so we can
estimate the timescale by neglecting the change of orbital period. The rate of
change of the spin energy is then given by the power injected into the spin $=T
\Omega_s$ so using Eq.~\ref{eq:torque}
\begin{equation}
I \Omega_s \dot{\Omega}_s = \frac{\Omega_s L_X}{\Omega_o - \Omega_s},
\end{equation}
where $I \sim (1/5) M_1 R_1^2$ is the moment-of-inertia of the primary star. 
Dividing through by $\Omega_o$ gives
\begin{equation}
\dot{\alpha} = \frac{L_X}{I \Omega_o^2} \frac{1}{1-\alpha}.
\end{equation}
This equation is identical to equation~8 of \cite{Wu:V407Vul} who also
find that $L_X \propto (1 - \alpha)^2$, so we can write
\begin{equation}
L_X = \frac{\left( L_X\right)_0}{(1-\alpha)^2_0} (1-\alpha)^2,
\end{equation}
where the subscript zeroes indicate initial values. Therefore
\begin{equation}
\dot{\alpha} = \frac{\left(L_X\right)_0}{I \Omega_o^2 (1-\alpha)_0^2} (1-\alpha),
\end{equation} 
which can be integrated to give
\begin{equation}
1 - \alpha = (1-\alpha)_0 \, e^{-t/\tau_s},
\end{equation} 
with the synchronisation timescale $\tau_s$ given by
\begin{equation}
\tau_s = \frac{I \Omega_o^2}{L_X} (1-\alpha)^2.
\end{equation}
Since
$L_X \propto (1-\alpha)^2$, the X-ray luminosity decays with a time constant of
$\tau_s/2$. Taking $M_1 = 0.6 \,\msun$, $R_1 = 0.01\,\rsun$, we find $\tau_s
\sim 4400 (1-\alpha)^2 (d/1\,\mathrm{kpc})^{-2}$ years for \vul, and $7 \times
10^5 (1-\alpha)^2 (d/1\,\mathrm{kpc})^{-2}$ years for \rxj. Putting in
\cite{Wu:V407Vul}'s value of $1-\alpha = 0.001$ leads to X-ray flux decay
constants of $0.8$~days for \vul\ and $4.2$~months for \rxj\ which are
incompatible with the $15$-odd years that these systems have been followed, let
alone plausible formation rates. Our timescale for \vul\ is a great deal shorter than the
$\sim 1000$ years quoted by \cite{Wu:V407Vul}.  It is not possible to determine from
\citeauthor{Wu:V407Vul} the exact cause of this difference, but we think that it 
is simply because their estimate was not specifically for \vul,
but applies to a lower luminosity model. Whatever the reason, the
important point is that the timescale is in fact much shorter than their
1000~year value, which was already hard to credit.  In
Fig.~\ref{fig:sync} we show numerical integrations of the synchronisation for
fixed dipole moments according to \cite{Wu:V407Vul}'s equations.
The initial conditions of the left- and right-hand panels were chosen to match
\vul\ and \rxj\ respectively. The integrations confirm that the UI model
can only sustain the observed X-ray fluxes for a very short time indeed.
In the integrations we account properly for the changing orbital period, and
hence in the right-hand panel a small residual asynchronism is seen at long
times as the spin frequency must always lag the orbital frequency by a small amount.

\subsection{Revision of the Unipolar Inductor model}
\label{sec:revision}

We have shown that \cite{Wu:V407Vul}'s model is ruled out by the
measured rates of period change in \vul\ and by its very short lifetime for both
\vul\ and \rxj. We now consider whether it can be adjusted to match them. The
problem stems from the inefficient X-ray generation as only 1 part in 1000 of
the power transferred from the orbit to the spin is dissipated. This can be
avoided if very different values of $\alpha = \Omega_s/\Omega_o$ are
considered. For \vul, a value of $|1-\alpha| \ga 100$ would reduce the induction
torque term to a value comparable to or smaller than the gravitational wave
term.  Since we expect prograde rotation ($\alpha > 0$), this means that $\alpha
\ga 100$, i.e. that the magnetic white dwarf must spin of order 100 times faster
than the orbit so that $P_\mathrm{spin} \la 6\,$sec, which is or order the
break-up spin rate of typical white dwarfs.  Since $\alpha > 1$, the
induction torque acts in the reverse sense to GR and therefore the large value
of $\alpha$ is needed not only to obtain the correct order-of-magnitude for
$\dot{P}$, but also to give the correct sign. 

Such a large value of $\alpha$ can only be attained through accretion-driven
spin-up.  The increased value of $1-\alpha$ also means that the magnetic moment
of the white dwarf must be much less than supposed by \cite{Wu:V407Vul}.
According to their model, the power dissipated in electric currents scales as
$\propto \mu^2 (1-\alpha)^2$, where $\mu$ is the magnetic moment of the white
dwarf.  Since in \vul\ we require $1-\alpha$ to be $10^5$ times larger than
\cite{Wu:V407Vul} assume, the magnetic moment of the white dwarf must
correspondingly drop by $10^5$, which would therefore end up at about $10^{27}$
to $10^{28}\,\mathrm{G}\,\mathrm{cm}^3$, corresponding to a surface field of
only 1 to $10\,\mathrm{G}$. We would then have a case of accretion onto a very
weakly magnetic white dwarf (non-magnetic as far as one could tell from
observations), which begs the question of why the accretion should ever have
stopped to allow the unipolar inductor to get going in \vul\ but not in similar
systems, such as the AM~CVn stars and cataclysmic variable stars. Occam's razor
suggests that if accretion has to be invoked to allow the UI model to work at
all, then one should favour accretion-only models, which we will look at in the
next section.

The much lower X-ray luminosity of \rxj\ leads to less stringent requirements.
For $1-\alpha = 0.001$ the induction term is only 10 times the GR term, and so a
modest increase leads to an acceptable match given the uncertainties. A
tougher constraint comes from the X-ray decay timescale, which at
4~months for $1-\alpha = 0.001$ is much too short. However, an increase to 
$1-\alpha = 0.01$, gives a time scale of $\sim 30$ years which is probably long
enough that it could not be ruled out by observations. 

The larger values of $1-\alpha$ needed to match \vul\ and \rxj\ have one further
consequence: they make it easier to see the phase shifts between optical and
X-ray pulses predicted under the UI model, but not observed, which were used by
\cite{Barros:V407Vul} to rule out the dipolar field geometry used by
\cite{Wu:V407Vul}. For small values of $1-\alpha$, it was always possible that 
an unlucky distribution of observations had lead to our missing these phase
shifts. The raised values of $1-\alpha$ and Fig.~6 of \cite{Barros:V407Vul} 
show that this is no longer the case.

In all the above, the X-ray luminosity, and therefore the distance, is crucial.
Smaller distances reduce $L_X$ and therefore alleviate the problems discussed
above.  We have assumed $d = 1\,\mathrm{kpc}$ in each case. For \vul\ this comes
from assuming that the variable is at the same distance as the G star which
dominates its spectrum (Steeghs et al., in prep.). For \rxj\ on the other hand,
the estimate comes from its blue colour, magnitude and adopting $M_V = 11$ as
the absolute magnitude of a white dwarf of comparable colour. Neither constraint
is secure, and lower distances are certainly possible. A crude lower limit comes
from the absence of detectable proper motion in either star, which suggests that $d >
100 \,\mathrm{pc}$. However, it is hard to see that \rxj\ can be near this limit
at the same time as being blue and faint ($U-B = -1.1$, $V = 21.1$
\citealt{Israel:RXJ0806}), unless the emitting area is much less than that of an
average white dwarf. A similar argument applies to \vul, because although it is
brighter than \rxj, much of its flux in $V$ comes from the G star, leaving the
variable comparable to \rxj\ in brightness. Thus $d > 300\,\mathrm{pc}$, is
probably a better lower limit, so the X-ray fluxes could perhaps be lowered by a
factor of 10. This would leave \rxj\ compatible with \cite{Wu:V407Vul}'s model, but \vul\
still a good way from it. Better constraints on the distances to these systems
are needed.

We now look at the whether accreting models can do any better in matching
the observed period decreases in \rxj\ and \vul.
 
\section{Period changes in accreting binary stars}

Assuming conservative mass transfer one can show that
\begin{equation}
\frac{\dot{P}}{P} = 3 \left( \frac{\dot{J}}{J} - (1-q)
\frac{\dot{M}_2}{M_2} \right), \label{eq:accrete}
\end{equation} 
where $q = M_2/M_1$ and $M_2$ is the mass of the donor. Since for
stability we require that $q < 1$, and since $\dot{J} < 0$ and $\dot{M}_2 < 0$, mass
transfer offsets the angular momentum loss term. The angular momentum
term depends upon whether the accretor's spin is strongly coupled to
the orbit or not. If it is we have simply $\dot{J} = \dot{J}_{GR}$;
if not then Eq.~1 from \cite{Marsh:mdot} gives
\begin{equation}
\dot{J} = \dot{J}_{GR} + \sqrt{G M_1 R_h} \dot{M}_2 ,
\end{equation}
where $R_h$ is the circularisation radius, which effectively increases the rate
of loss of orbital angular momentum.

An illuminating comparison with the section~\ref{sec:induction} can be made by
calculating the ``effective'' angular momentum loss rate
$\dot{J}_{eff}$ such that the detached formula can still be applied as $\dot{P}/P = 3
\dot{J}_{eff}/J$ because this can then be compared directly to Eq.~\ref{eq:jdot}. 
Assuming that $L_X = - G M_1 \dot{M}_2 / R_1$ where $M_1$ and $R_1$
are the mass and radius of the accretor, and using
Eq.~\ref{eq:accrete} (strong coupling), one can show that 
\begin{equation}
\dot{J}_{eff} = \dot{J}_{GR} + (1-q) \frac{R_1}{a}
\frac{L_X}{\Omega_o}, \label{eq:accrete1}
\end{equation}
in the strongly coupled case where $a$ is the orbital separation.
Comparing with Eq.~\ref{eq:jdot}, we see that the factor
$1/(1-\alpha) \sim 1000$, which causes \cite{Wu:V407Vul}'s model so much
trouble, is replaced by $(1 - q) R_1 / a \sim 0.1$. This is a considerable
improvement, and given the uncertainties in $L_X$ and $L_{GR}$
discussed in section~\ref{sec:revision}, could be made consistent with
the observed $\dot{P}$ values for both systems. 

For \rxj, the X-ray luminosity is already lower than the gravitational wave
luminosity, as pointed out by \cite{Strohmayer:RXJ0806_2}, and so
Eq.~\ref{eq:accrete1} predicts that the observed period change must be close to
the pure GR one. Weak coupling is better still since it increases the loss of
angular momentum from the orbit, but since it is otherwise qualitatively the
same, we won't discuss it further.

This discussion hides the ugly truth about accreting models which is that the
\emph{equilibrium} value of $\mtwodot$ leads to an \emph{increasing} period. Put
differently, the current mass transfer rates in \vul\ and \rxj\ must be $\la
60$\% of their equilibrium values in order for their periods to decrease at all
\citep{Marsh:mdot} (with some uncertainty over \vul\ since some of its apparent
period change could be caused by light-travel time effects if it is truly
associated with the G star mentioned earlier). This is the price one must pay to
accept accreting models. We regard it as a small one because there are already
examples of systems which do not have mass transfer rates that match
expectations, such as long period dwarf novae, and there are ways to make the
mass transfer rate deviate from its equilibrium value as we will discuss below.

If the mass transfer rate is below its equilibrium value, how long will it stay
like this and would we expect to see a noticeable increase over the course of a
few years? The mass transfer rate in the two systems is expected to be in the
adiabatic regime \citep{Webbink:DDs,Marsh:mdot} for which $\dot{M} \propto
\Delta^3$ where $\Delta = R_2 - R_L$ is the amount by which the donor overfills
its Roche lobe \citep{Marsh:mdot}. Therefore the rate of change of mass transfer
rate is given by
\begin{equation}
  \ddot{M} = 3 \frac{\dot{M}}{\Delta} \dot{\Delta},
\end{equation}
and so the
timescale for significant change of the mass transfer rate is given by
\begin{equation}
\tau_c = \frac{ \dot{M} }{\ddot{M}} = \frac{1}{3} \frac{\Delta}{\dot{\Delta}}.
\end{equation}
Using Equation~19 of \citep{Marsh:mdot}, assuming strong coupling and neglecting
the $\dot{M}$ dependent term in order to calculate the maximum rate of change
of $\dot{M}$ gives
\begin{equation}
\tau_c \approx = \frac{1}{6} \frac{\Delta}{R_2} \left(
  \frac{J_{orb}}{-\dot{J}_{GR}} \right) .
\end{equation}
Taking $M_1 = 0.6\,\msun$, and $M_2 = 0.07$ and $0.12\,\msun$ for \vul\ and
\rxj\ respectively, and using equations 10-12 of \citep{Marsh:mdot} to calculate
$\Delta$, we find timescales of $\sim 1000$ and $100$ years for significant
alterations in the mass transfer rate to occur in \vul\ and \rxj\ respectively.
These numbers suggest that below-equilibrium transfer rates can be sustained for
many years, as observed. We must appeal to some unknown mechanism to force the
departure from equilibrium in the first place, but there is no shortage of
candidates. For example, a star-spot moving over the inner Lagrangian point is
one possibility \citep{Livio:starspots,Hessman:AMHer,Hessman:dynamo},
irradiation-induced cycles another \citep{Ritter:cycle}, widening of the orbit
caused by the mass ejected in nova explosions a third \citep{Shara:hibernate}
and synchronisation-induced detachment a fourth \citep{Lamb:synchronise}. One
can even circumvent the relation between mass transfer rate and period
derivative altogether through alterations in the structures of the stars
\citep{Applegate:pdot}, a mechanism which is thought to be responsible for
cyclical variations in the periods of many eclipsing cataclysmic variables
\citep{Baptista:pdot}. This mechanism can make the period decrease even if
mass transfer occurs at the equilibrium rate, although the low X-ray luminosity
of \rxj\ means that there remains a need for below-equilibrium mass transfer
whatever the cause of the period decrease. Which of these mechanisms, if any,
can be applied to double white dwarfs is not clear, but neither is it clear that
any of them are \emph{not} applicable.

\section{Conclusions}

We have shown that, contrary to widespread assumption, the unipolar inductor
model \citep{Wu:V407Vul} does a poor job at predicting the magnitude of the
period changes observed in the two candidate ultra-compact binary stars \vul\
and \rxj\ in the sense that the predicted changes are much larger than those
observed. This removes the main piece of evidence supporting the model. The
reason for this is that if there is only a small asynchronism between the spin
and orbital periods, then much more energy is transferred between the orbit and spin than
is dissipated in X-rays. The problem is particularly acute in the case of \vul\
for which the unipolar inductor model predicts a rate of period change some $100$,$000$
times greater than is observed. The unipolar inductor model can only apply to \vul\
if its primary star rotates much faster than the 570 second putative orbital
period; a larger asynchronism is also required to allow the unipolar inductor
model to last more than a few years. This suggests that for the unipolar
inductor model to work, accretion is a necessary precursor, without there being
any obvious reason for it to have ceased. The problem with accretion-only models
remains that the equilibrium mass transfer rates should lead to increasing
orbital periods (as opposed to the observed decreases), and so the two systems
must currently be transferring mass at below equilibrium rates or other
mechanisms must be affecting the orbital periods if accretion is
to work. Given that there are many examples of systems where this is the case,
this difficulty seems less significant than those facing the unipolar inductor
model. Nevertheless, the nature of these systems remains far from clear, and
observational efforts to pin them down must continue to be pursued.

\section*{Acknowledgements}
TRM acknowledges the financial support of a PPARC SRF. 
GN is supported by NWO-VENI grant 639.041.405.
This research has made use of NASA's Astrophysics Data System 
Bibliographic Services and the SIMBAD database,
operated at CDS, Strasbourg, France.

\bibliographystyle{mn_new}

\bibliography{refs}

\label{lastpage}

\end{document}